\documentclass[iop]{emulateapj}
\usepackage{natbib}
\usepackage{apjfonts}
\usepackage{epsfig}

\newcommand{\etal}{et al.}

\def\K{{\rm\thinspace K}}

\def\kpc{{\rm\thinspace kpc}}

\def\Mpc{{\rm\thinspace Mpc}}

\def\Msun{\hbox{$\rm\thinspace M_{\odot}$}}

\def\pc{{\rm\thinspace pc}}

\def\Mdot{\hbox{$\dot M$}}

\def\<{\thinspace}
\def\Mpc{{\rm\thinspace Mpc}}
\def\Gpc{{\rm\thinspace Gpc}}

\def\msun{{\rm M_{\odot}}}

\def\simlt{\lower.5ex\hbox{$\; \buildrel < \over \sim \;$}}
\def\lesssim{\lower.5ex\hbox{$\; \buildrel < \over \sim \;$}}
\def\simgt{\lower.5ex\hbox{$\; \buildrel > \over \sim \;$}}

\def\h{{\rm\thinspace h}}
\newcommand\beq{\begin{equation}}
\newcommand\eeq{\end{equation}}
\newcommand\beqa{\begin{eqnarray}}
\newcommand\eeqa{\end{eqnarray}}

\shorttitle{Cold flows and the first quasars} \shortauthors{Di Matteo \etal}
\slugcomment{Submitted to ApJ 07/05/11}
\begin{document}

\title{Cold gas flows and the first quasars in cosmological simulations}

\author{T. Di Matteo,\altaffilmark{1},
  N. Khandai \altaffilmark{1}, C. DeGraf,\altaffilmark{1}
 Y. Feng \altaffilmark{1}, R.A.C. Croft \altaffilmark{1}, 
J. Lopez \altaffilmark{2}, V. Springel \altaffilmark{3,4}} 

\altaffiltext{1} {McWilliams Center for Cosmology, Carnegie
  Mellon University, 5000 Forbes Avenue, Pittsburgh, PA 15213}

\altaffiltext{2} {Computer Science Department, Carnegie, Mellon University,
5000 Forbes Avenue, Pittsburgh, PA 15213} 

\altaffiltext{3}{Heidelberg Institute for Theoretical Studies
Schloss-Wolfsbrunnenweg 35, 68118 Heidelberg, Germany}

\altaffiltext{4} {Zentrum f\"{u}r Astronomie der
Universit\"{a}t Heidelberg, Astronomisches
Recheninstitut, M\"{o}nchhofstr. 12-14, 69120 Heidelberg, Germany}

\begin{abstract}
  Observations of the most distant bright quasars imply that billion solar
  mass supermassive black holes (SMBH) have to be assembled within the first
  eight hundred million years. Under our standard galaxy formation scenario
  such fast growth implies large gas densities providing sustained accretion
  at critical or supercritical rates onto an initial black hole seed. It has
  been a long standing question whether and how such high black hole accretion
  rates can be achieved and sustained at the centers of early galaxies.  Here
  we use our new {\it MassiveBlack} cosmological hydrodynamic simulation
  covering a volume $(0.75 \Gpc)^3$ appropriate for studying the rare first
  quasars to show that steady high density cold gas flows responsible for
  assembling the first galaxies produce the high gas densities that lead to
  sustained critical accretion rates and hence rapid growth commensurate with
  the existence of $\sim 10^9 \Msun$ black holes as early as $z\sim 7$. We
  find that under these conditions quasar feedback is not effective at
  stopping the cold gas from penetrating the central regions and hence cannot
  quench the accretion until the host galaxy reaches $M_{\rm halo} \simgt
  10^{12} \Msun$. This cold-flow driven scenario for the formation of quasars
  implies that they should be ubiquitous in galaxies in the early universe and
  that major (proto)galaxy mergers are not a requirement for efficient fuel supply
  and growth, particularly for the earliest SMBHs.
\end{abstract}


\section{Introduction\label{sec:intro}}
It is now well established that the properties of supermassive black holes
(SMBH) found at the centers of galaxies today are tightly coupled to those of
their hosts implying a strong link between black hole and galaxy
formation. The strongest direct constraint on the high-redshift evolution of
SMBHs comes from the observations of the luminous quasars at $z \sim 6$ in the
Sloan Digital Sky Survey (SDSS) \citep{Fan2006, Jiang2009} and even more
recently at $z=7$ \citep{Mortlock2011}.  Although rare (the comoving space
density of $z \sim 6$ quasars is roughly $n \sim$ a few$ \Gpc^{-3}$) the
inferred hole masses of these quasars are in excess of $10^9~\Msun$ comparable
to the masses of the most massive black holes in the Universe today.
The origin of these massive black hole seed and the physical conditions that
allow early growth to supermassive black holes remain a challenging
problem. 

\begin{figure*}
\centering
\hbox{
\hspace{-0.1cm}
\includegraphics[scale=1.02]{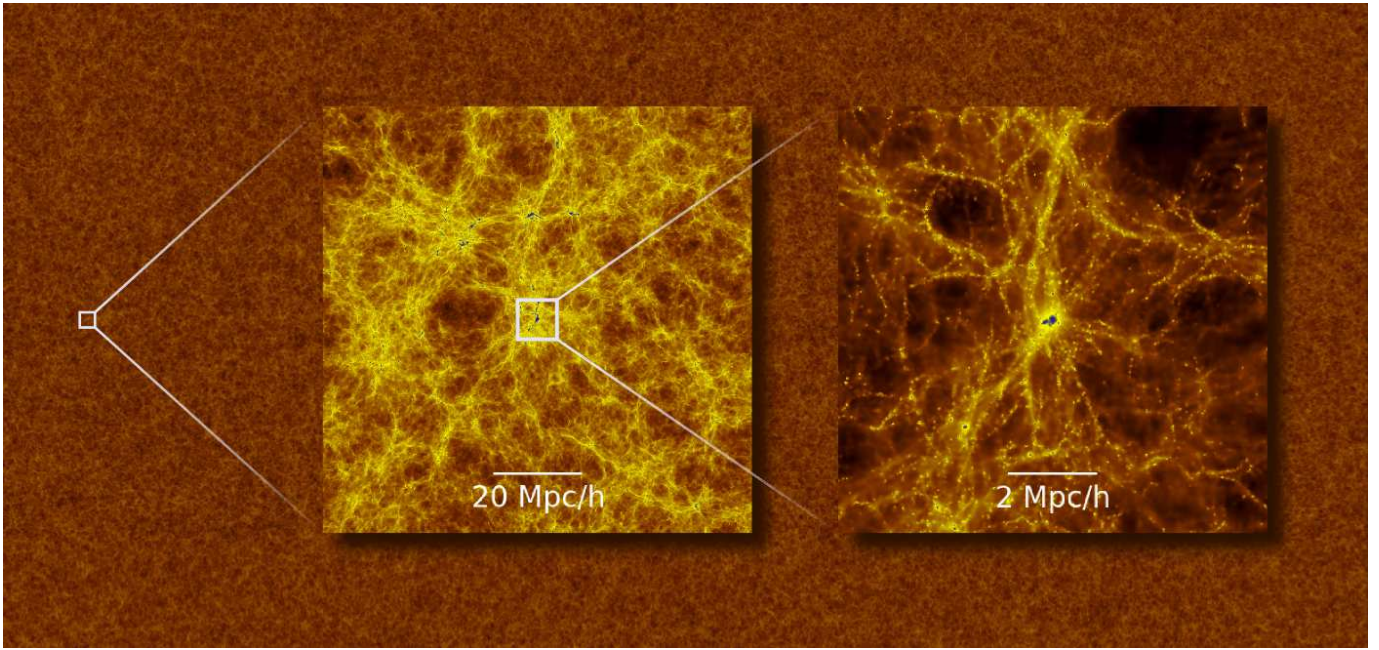}}
\label{fullsim}
\caption{The cosmological mass distribution in our the simulation volume at
  $z=5$. The projected gas density over the whole volume ('unwrapped' into
  2D) is shown in the large scale image. The two overlaid panels show
  successive zoom-ins by factor of 10, center on the region where the most massive
  black hole is found.}
\end{figure*}
In order to have had sufficient time to build up via gas accretion and BH
mergers (resulting from the hierarchical merging of their host halos) the
first 'seed' black holes must have appeared at early epoch, $z > 10$.  The
origin and nature of this seed population remain uncertain. Two distinct
populations of seed masses, in the range of $100-10^5~\msun$ have
been proposed: the small mass seeds are usually thought to be the remnants of
the first generation of PopIII stars formed of metal-free gas at $z\sim
20-30$ \citep[e.g.]{Bromm1999, Abel2000, Nakamura2001, Yoshida2003, Gao2006},
while the large seeds form in direct dynamical collapse in metal-free
galaxies \citep[although see also Mayer et al. 2010 for direct collapse into a
massive blackhole in metal enriched regime]{Koushiappas2004, Begelman2006}.

Growing the seeds to $~10^9 \Msun$ in less than a billion years requires
extremely large accretion rates - as mergers between black holes are too rare and
too inefficient for significant growth.  For a black hole accreting at the
critical Eddington accretion rate its luminosity $L_{\rm Edd} = (4 \pi G c
m_p)/ \sigma_{T} M_{BH} = \eta \Mdot_{\rm Edd} c^2$ (where $G$, $c$, $m_{p}$
and $\sigma_{T}$ are the gravitational constant, speed of light, proton mass
and Thomson cross section, and $\eta$ is the standard accretion efficiency)
implies an exponential growth at the characteristic Eddington timescales,
$t_{\rm Edd} = 450 \eta / (1- \eta)$ Myr, such that $M_{\rm BH} = M_{\rm seed}
e^{t/t_{\rm Edd}}$. For a seed mass ranging from $M_{\rm seed} \sim 100-10^5
\Msun$ this requires $10-17$ e-foldings to reach $M_{\rm BH} \sim 10^9
\Msun$. The crucial question (for any given seed model) is then where (if at
all, in which kind of halos) and how (at what gas inflow rates) such vigorous
accretion can be sustained at these early times.

As bright quasars are likely to occur in extremely rare high-density peaks in
the early universe, large computational volumes are needed to study them. Here
we use a new large cosmological Smooth Particle hydrodynamics (SPH)
simulation, {\it Massive Black} (covering a volume of $[0.75 \Gpc]^{3}$) with
sufficiently high resolution (over 65 billion particles) to be able to include
tested prescriptions for star formation, black hole accretion and associated
feedback processes to investigate whether and if such objects may be formed
within our standard structure formation models.  Crucially, our {\it
  MassiveBlack} simulation is of sufficiently high-resolution to allow us to
follow the mass distribution in the inner regions of galaxies and hence model
star formation and black hole growth directly and self-consistently while
still evolving a close to Gigaparsec scale region. It therefore provides a
unique framework to study the formation of the first quasars.


\section{Methodology\label{sec:method}}
\subsection{Simulation run: ``Massive Black''}
Our new simulation has been performed with the cosmological TreePM-Smooth Particle
Hydrodynamics (SPH) code {\small P-GADGET}, a {\it hybrid} version of the
parallel code {\small GADGET2}~\citep{Springel2005d} which has been
extensively modified and upgraded 
to run on the new generation of Petaflop scale supercomputers
(e.g. machines like the upcoming BlueWaters at NCSA). The major improvement
over previous versions of {\small GADGET} is in the use of threads in both the
gravity and SPH part of the code which allows the effective use of multi core
(currently 8-12 cores per node) processors combined with an optimum number of
MPI task per node. Here we present initial results from the largest
cosmological smooth particle hydrodynamic simulation to date which was run on $10^5$
cores corresponding to the entire Cray-XT5 ``Kraken'' at NICS.  The {\it
  MassiveBlack} simulation contains $N_{\it part} = 2 \times 3200^3 = 65.5$
billion particles in a volume of $533 \Mpc /h$ on a side with a gravitational
smoothing length $\epsilon = 5.5 \kpc/h$ in comoving units). The gas and dark
matter particle masses are $m_{\rm g} = 5.7 \times 10^7 \Msun$ and $m_{\rm DM}
= 2.8 \times 10^8 \Msun$ respectively. This run contains gravity and
hydrodynamics but also extra physics (subgrid modeling) for star
formation \citep{Springel2003a}, black holes and associated feedback
processes. The simulation has currently been run from $z=159$ to $z=4.75$
(beyond our original target redshift of $z=6$). For this massive calculation
it is currently prohibitive to push it to $z=0$ as this would require an
unreasonable amount of computational time on the world's current fastest
supercomputers. The simulated redshift range probes early structure formation
and the emergence of the first galaxies and quasars.

\subsection{Black hole Accretion and Feedback Model}
The prescription for accretion and associated feedback from massive black
holes has been developed by ~\cite{DiMatteo2005, Springel2005a}.  Detailed
studies of this implementation in cosmological simulations and associated
predictions \citep{Sijacki2007, Li2007, DiMatteo2008, Croft2009, Sijacki2009,
  Colberg2008, DeGraf2010, DeGraf2011, Booth2011} have shown that it can reproduce all the
basic properties of black hole growth, the observed $M_{BH}-\sigma$,
relation \citep{DiMatteo2008}, the quasar luminosity function
\citep{DeGraf2010} and its evolution as well as the spatial clustering of
quasars \citep{DeGraf2011a}. In a nutshell our black hole accretion
and feedback model \citep{DiMatteo2008} consists of representing black holes by
collisionless particles that grow in mass (from an initial seed black hole) by
accreting gas in their environments. A fraction of the radiative energy
released by the accreted material is assumed to couple thermally to nearby gas
and influence its motion and thermodynamic state (typically referred to as BH
feedback). Our underlying assumption is that the large-scale feeding of
galactic nuclei with gas (which is resolved in our simulations) is ultimately
the critical process that determines the growth of massive black holes and the
peak of the quasar phase \citep{DiMatteo2008, DeGraf2010, DeGraf2011}. The
model, therefore needs to be viewed in the context of cosmological growth of
black holes and not detailed accretion physics. While a more detailed
treatment of this is certainly desirable and begins to be possible for
individual galaxies \citep[e.g.]{Kim2011, HopkinsQuataert2010}
it is still infeasible for cosmological simulations that seek to follow whole
populations of galaxies and their BHs.


\begin{figure*}[t]
\centering
\includegraphics[scale=1]{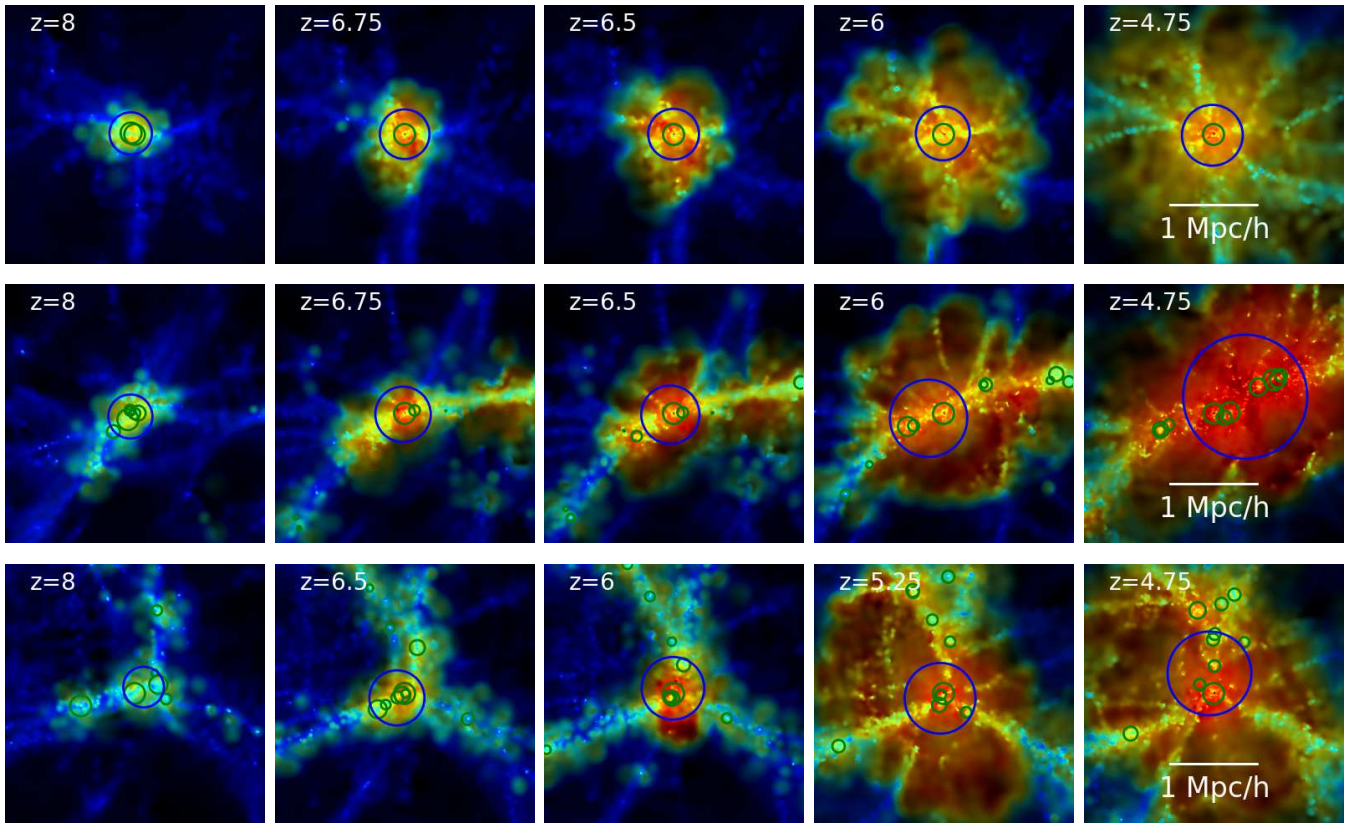}
\label{first_qso}
\caption{Snapshots of the evolution of the cold-flow-fed massive black
  holes. The images visualise the projected gas distribution color coded by
  temperature around three example quasars (one in each row) across five
  different redshifts (labelled on each of the five panels; left to right).
  The projected density ranges from $\sim 10^{-2}$ to $\sim 10^2 \h \Msun
  \pc^{-2}$ and the temperature from $\sim 10^4$ (blue colors) to $10^8 \K$
  (red colors). The quasar positions are indicated by the green circles and
  the virial radius of each halo by the blue circles. The images show the
  typical structure of cold streams that penetrate the halo all the way into
  the central regions of galaxies.  At first ($z \simgt 7.5$) the gas is cold
  and the black hole is still of relatively low mass, but then below this
  redshift the black hole growth exponentiates as increasing amounts of cold
  gas is fed into the central regions. Black hole feedback heats the gas, but
  does not distrupt the cold streams until $ z \simlt 6$.}
\end{figure*}

We introduce collisionless `sink' particles in the simulations to model black
holes at the centers of forming minihalos. In order to achieve this we keep
track of the formation of minihalos by running a friends-of-friends (FOF)
group finder on the fly.  The group finder is run on sufficiently closely
spaced intervals to identify the newly collapsing halos in which we place a
black hole seed of fixed mass, $ M = 10^5 h^{-1}$\,M$_\odot$ (if they do not
already contain a BH). In practice most of the halos of this mass and their
black holes are formed between $z=15$ and $z=30$.  The black hole particle
then grows in mass via accretion of surrounding gas according to $\dot{M}_{\rm
  BH} = \frac {4 \pi G^2 M_{\rm BH}^2 \rho}{(c_s^2 + v^2)^{3/2}}$ (where
$\rho$ and $c_s$ are the density and sound speed of the hot and cold phase of
the ISM gas which when taken into account appropriately as in
\citet{Pelupessy2007} - this eliminates the need for a correction factor
$\alpha$ previously introduced - and $v_{\rm BH}$ is the velocity of the black
hole relative to the gas) and by merging with other black holes.  We limit the
accretion rate to Eddington (or a few times Eddington). A similar Bondi model
has also been used by \cite{Johnsonbromm2007} to study the growth of the first
massive PopII black holes remnants. 

The radiated luminosity, $L_{\rm r}$, from
the black hole is related to the 
accretion rate, $\dot {M}_{\rm BH}$ as $ {{L_{\rm
      r}}= \epsilon_{\rm r} \, ({\dot {M}_{\rm BH} \times c^2}}) \,$, where we
take the standard mean value $\epsilon_{\rm r} =0.1$. Some coupling between
the liberated luminosity and the surrounding gas is expected: in the
simulation 5\% of the luminosity is (isotropically) deposited as thermal
energy in the local black hole kernel, providing some form of feedback energy
\citep{DiMatteo2005}. This model of AGN feedback as isotropic thermal coupling
to the surrounding gas, albeit simple, is a reasonable approximation to any
physical mechanism which leads to a shock front which isotropizes and becomes
well mixed over physical scales smaller than those relevant in our simulations
and on timescales smaller than the dynamical timescales of the
halos \citep{DiMatteo2008, HopkinsHernquist2006}. Two black hole particles
merge if they come within the spatial resolution (i.e. within the local SPH
smoothing length) with relative speed below the local sound speed.
\begin{figure*}[ht]
\centering
\hbox{
\hspace{-0.5cm}
\includegraphics[scale=0.55]{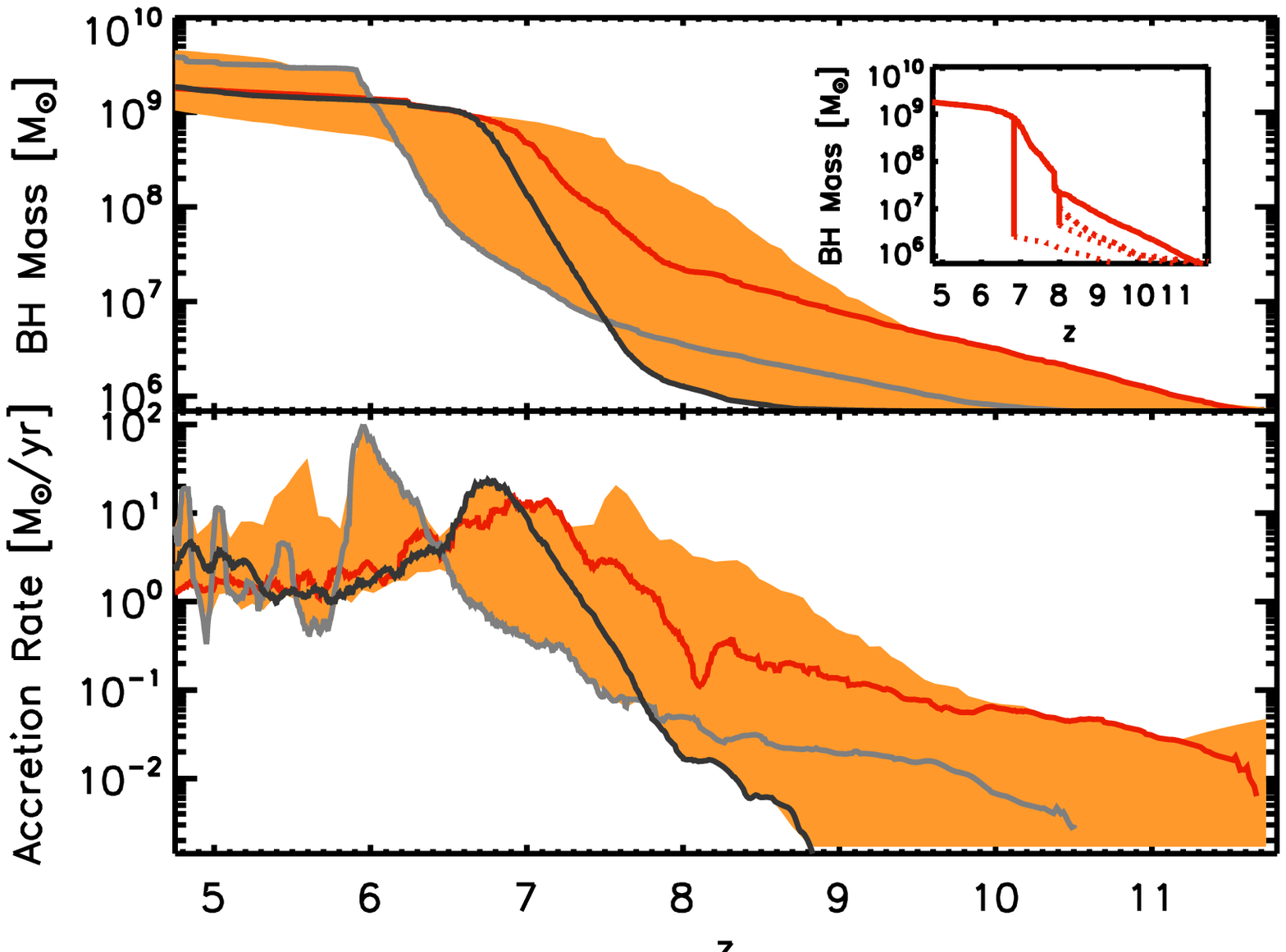}
\hspace{-0.8cm}
\includegraphics[scale=0.55]{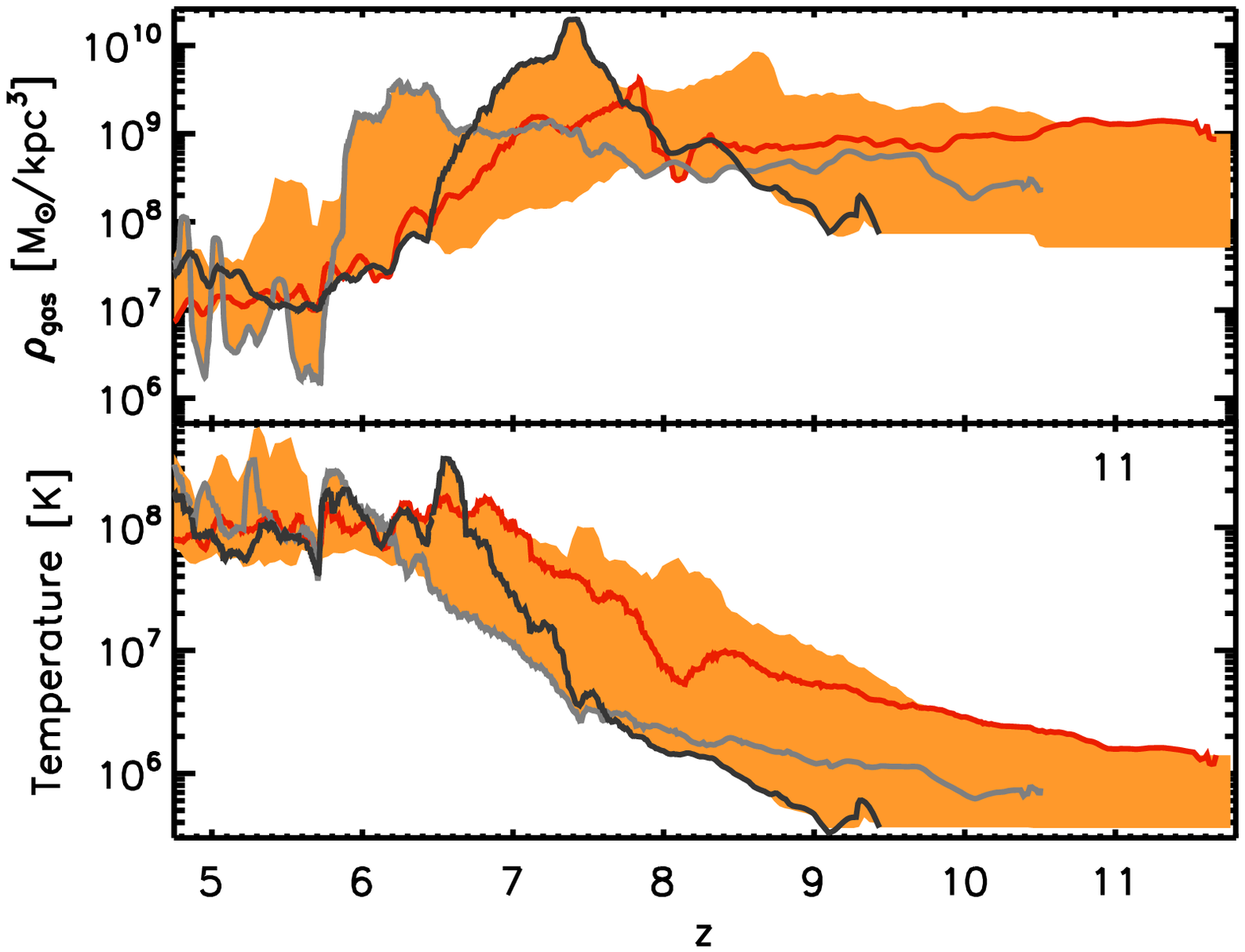}}
\label{massmdot_qso}
\caption{The black hole mass and the black hole accretion rate versus redshift
  are shown in the top and bottom panel on the left. The corresponding gas
  densities and temperatures of the accreting gas are shown in the top and
  bottom panels on the right. The lines represent the three black holes shown
  in Fig. 2 and the shaded orange band indicates the full range of properties
  encompassed by our most massive black hole sample.}
\end{figure*}

We note that at least two independent groups\citep{Booth2009, Johansson2008}
now have also adopted the same modeling for black hole accretion, feedback and
BH mergers in the context of hydrodynamic simulations. These independent
works, and in particular, the cosmological simulations by \citep{Booth2009}
(part of the OWL program) have allowed to independently explore the parameter
space of the reference model of \citep{DiMatteo2008}, as well as variations of
our model prescriptions. This large body of already existing work and
investigations make this particular model a good choice for more detailed
studies of the growth of the first quasars which is the subject we focus on
here. In associated publications we show that the {\it MassiveBlack} quasars
are fully consistent with all the fundamental statistical constraints for the
observed populations of high redshift quasars, and in particular with the
observed luminosity functions of quasars and the high redshift clustering
\citep{DeGraf2011b} and basic properties of quasar hosts \citep{khandai2011}.
To produce the black hole history from the simulation data we rely on SQL
databases developed by \citet{Lopez2011}.

\section{Results}
The cosmological gas density distribution in the full volume of the
{\it MassiveBlack} is shown in the large scale image of Figure~1. The large
panel shows the whole of the 3D simulation volume ``unwrapped'' into a 2D
image slice \citep{Feng11} at $z=5$.  At these large scales the density
distribution of the universe appears fairly uniform. The resolution however is
sufficiently fine to make it possible to zoom into increasingly smaller
regions and search for massive black holes that have experienced significant
growth. Superimposed onto the full scale image we show a zoomed region (scale
of $100 \Mpc\; h^{-1}$ and $10 \Mpc \;h^{-1}$ on a side from left to right, 
respectively) around one of the largest black holes/quasars at this time. On
these scales the images show the typical filaments that compose the cosmic web
and in particular how the first massive quasars (the most massive black hole,
at the center has a mass $M_{\rm BH} \sim 3 \times 10^9 \Msun$) form at the same
type of intersection/nodes of filaments that are the expected locations of
rare (massive) dark matter halos. Remarkably we do find ten black holes in the
volume that have grown to about a billion solar masses by $z \sim 6$ or
earlier (and many other of smaller masses). 

Figure 2 shows the environment and its evolution (five time-lapses from about
$z \sim 9$ to $z\sim 5$) of three examples from within this sample. Their
detailed mass assembly history is shown in Fig.~3. The panels in Figure~2 show
the evolution of the gas density color coded by temperature. These objects are
found to be continuously fed by intense streams of high density gas
(consistent with the cold-accretion picture for the growth of galaxies at
intermediate/high redshift by \cite[e.g.]{DekelBirnboin2006}. During these times
Eddington accretion is attained and sustained. By showing the temperature of
the gas, the images clearly make visible an expanding 'bubble' (emerging from
about $z\sim 6.5-7$) of hot gas (red colors) around the central quasars (whose
positions are indicated by green circles). This bubble created by the BH
feedback is more or less confined within the halo (the virial radius of the
halo is shown by the blue circles in Fig.~2) for $ z \simlt 6$. Below this
redshift, and rather abruptly, the energy released by the quasars heats and
expels the gas as a wind well beyond the halo. Black hole growth has now
become self-regulated (see also Figure 3, Eddington rates are reached only
sporadically). Although the effects of quasar feedback in our model have been
studied in detail previously \citep{DiMatteo2005, DiMatteo2008} what is
remarkable here is that even though feedback energy consistently heats the gas
within and eventually beyond the scale of halos, it does not do much to the
streams of in-flowing cold gas. There the gas density is so high (e.g. Fig.~3)
that the gas cannot be stopped because, it is too difficult to couple
enough feedback energy to it to disrupt the flow. The streams get somewhat
(albeit not completely) disrupted only at $z\simlt 6$.  Before this happens
billion solar mass black holes are already assembled (see Fig.~3), a process
that takes a few hundred million years.

Sustained phases of Eddington accretion onto these black holes start as early
as $z \sim 9-10$ and go on uninterrupted until $z \sim 6-7$, leading to BH
masses of the order of $10^9 \Msun$ in the first massive halos of roughly
$M_{\rm halo}\sim 10^{12} \Msun$ at $z \sim 6 -7$ (rare $3-4 \sigma$ peaks of
the density distribution). We have indeed a few objects (the range of the 10
most massive is shown by the orange areas in Fig.~3) that reach these high
masses already at $z\sim7$ consistent with
\citep{Mortlock2011}. Interestingly, this process of BH assembly is apparently
facilited by the ``cold flows'' picture of galaxy formation which has
revolutionized our understanding of galaxy assembly below the threshold dark
matter mass of $M_{\rm halo}\sim 10^{12} \Msun$ \citep{DekelBirnboin2006,
  Dekel2009, Keres2005, Keres2009}.  With {\it MassiveBlack} we are able to
trace the formation of the first, rare massive halos, those which are mostly
assembled by high density (high redshift) cold streams \citep{Dekel2009}.  We
find that these same streams easily penetrate all the way into central regions
of galaxies even in the presence of strong feedback. We find however that once
$M_{\rm halo}\sim 10^{12} \Msun$ and halos enter the regime where they are
shock-heated \citep{Dekel2009} the BH growth becomes finally self-regulated
(see accretion rate evolution in Fig.~3). The temperature of the accreting gas
also is raised well above the virial temperature, $T_{\rm vir} \sim 10^7 \K$,
rendering some of the gas unbound; Fig.~3). This the point the black hole
masses level off at a few $10^9 \Msun$ (Fig.~3). Even though at $z < 6$ in
massive halos the outflows clearly affect the incoming gas it is not clear
they fully distrupt the cold flows. According to
\citet[e.g]{DekelBirnboin2006}, albeit based on somewhat lower redshift than
what we probe, cold streams should still not be fully distrupted in this rare
massive objects even if a dilute hot medium forms. In future work we will
address this issue in our simulations in some more details.

We note that the black holes occupying these rare massive halos hardly undergo
any mergers, indicative that their hosts are not undergoing major galaxies
mergers. In the inset in Fig. 3 we show the histories of all black holes that
merge into the main progenitor, which is typical of what we see: a few and
typically minor mergers occur.  At the redshifts relevant for the first
quasars, however, galaxy formation is very different: mergers are still
extremely rare events and halos have not yet assembled above their shock
heating scale \citep{Dekel2009}, so that quasars are fueled directly by cold
streams/flows.

\begin{figure}
\centering
\hbox{
\hspace{-0.5cm}
\includegraphics[scale=0.35]{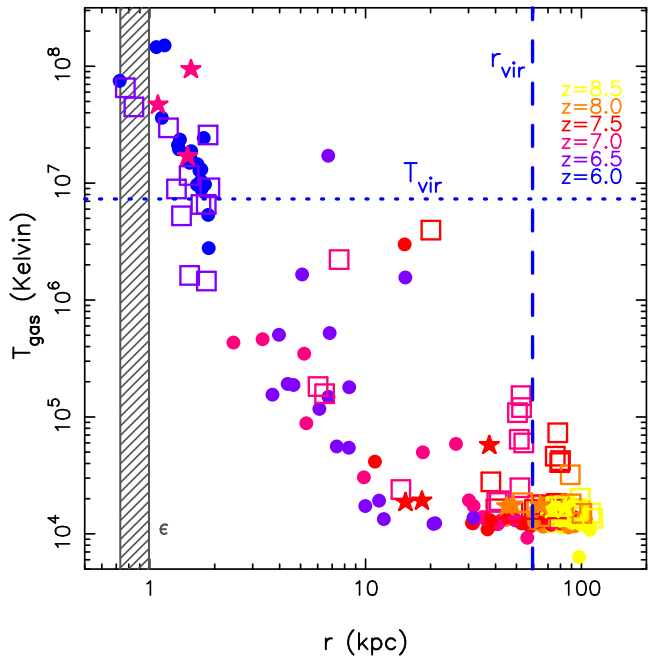}}
\label{coldflow_qso}
\caption{The temperature of the gas that is eventually accreted onto the black
  hole is shown as a function of its position within the host halo ($r$ is
  plotted in physical units). The plot shows a typical example for one of the
  host halos in Fig. 3. The points with different symbols representing the
  groups of 32 nearest neighbours at different redshifts (in the colors
  indicated by the legend) all traced back to $z=8.5$. The gray hatched region
  show the gravitational softening lengths}.
\end{figure}

To illustrate directly the origin of the gas fueling the first quasars we
track in the simulation the prior history of particles that end up in the
vicinity of the black hole, and hence contribute to the accretion. Figure~4
shows one example of the temperature as a function of radius of particles that
have participated into the accretion onto the BH between $z=6$ and $z=8.5$
(within the smoothing length of the black hole particle).  The plot shows that
particles always remain cold as they enter the virial radius well
into the ten of $\kpc$ region in-fact all the way
into the region of from which accretion onto the black hole occurs. Here some
of the gas is heated, as expected, by BH feedback. The temperature of the gas
feeding the black hole is fully commensurate with that of cold flows and always
remains below $T_{\rm gas} \sim 10^{5.5-6} \K$ just as predicted in galaxy
formation \cite{DekelBirnboin2006, Keres2005}. In the shock-heated regime the
temperature of the gas would rise to $T_{\rm vir}$ as it enters the
halo. However this is not seen.

\section{Conclusions}
With our new large cosmological simulation {\it MassiveBlack} we show that the
short timescale associated with infall via cold flows and the short cooling
timescales in cold radial streams that penetrate the halo render the flow into
the central regions unstoppable by feedback allowing it to easily sustain BH
growth at the Eddington rates to build up the required BH masses by $z=6-7$.
One consequence of this scenario is that BH masses at these redshifts are
expected to show deviations from the local $M_{BH} - \sigma$ relation. BH
masses assembled faster (most growth occurs over few hundred million years)
than the stellar spheroid (assembled over Gyrs timescales).  Black hole masses
larger than that inferred from the local $M_{BH} - \sigma$ relation have
infact been suggested for the first quasars.  Stream fed accretion will still
be relevant over the peak of the quasar phase but mergers (as merger rates
peak closer to those redshift) will become an increasingly major player in
their growth and formation. We speculate however that most of the growth of a
quasar's mass is likely to always occur before the shock heating scale of an
halo is reached much like most of its star formation rate
\citep{Dekel2009}. We will investigate this further in our large volume in
future work.  Our scenario is somewhat similar to that proposed by
\citet{Mayer2010} or \citet{Li2007} yet, crucially it does not rely on a major
merger to induce the strong gas inflows but points to a more common origin for
them particularly at these redshifts (which we could find by virtue of having
a large volume in our simulation, see also \citet{Sijacki2009} who followed the
build-up of a single hihigh-z quasar). As we have shown, relaxing the
constraint for massive mergers makes it plausible to attain quite commonly
large black hole masses as high as $z=7$ commensurate with \citet{Mortlock2011}.

\acknowledgments Stimulating discussions at the Aspen Center for Physics and in
  particular with Eliot Quataert, Romain Teyssier, Lucio Mayer and Elena Rossi
  are greatly aknowledged. We partiuclarly thank A. Dekel for comments on the manuscript.
  This research has been supported
  by the National Science Foundation (NSF) grant AST 1009781 and NSF OCI
  0749212. Computations were performed on the Cray XT5 supercomputer
  ``Kraken'' at National Institute for Computational Sciences (NICS). Imaging
  and analysis were carried out on the SGI UV ``Blacklight'' at the Pittsburgh
  Supercomputer Center (PSC) both part of the NSF Teragrid Cyberinfrastructure
  and on facilities provided by the Moore Foundation at CMU. The simulations
  are uploaded on GIGAPAN and can be viewed with black holes on {\it
  http://www.gigapan.org/gigapans/76215/}.
\bibliographystyle{apj}
\bibliography{ms}

\end{document}